\documentclass{article}
\usepackage[utf8]{inputenc}
\usepackage{lineno}
\usepackage{authblk}
\usepackage{natbib}
\usepackage{graphicx}
\usepackage{hyperref}
\usepackage{pbox}
\usepackage{dsfont}
\usepackage{amsmath}
\usepackage[table]{xcolor}
\usepackage{multirow}
\usepackage{color, soul}
\usepackage[strings]{underscore}

\title{Detecting seasonal episodic-like spatiotemporal memory patterns using animal movement modelling}
\author[1,*]{Peter R. Thompson}
\author[1]{Andrew E. Derocher}
\author[2,3]{Mark A. Edwards}
\author[1,4]{Mark A. Lewis}
\affil[1]{Department of Biological Sciences, University of Alberta, Edmonton, AB, Canada}
\affil[2]{Mammalogy Department, Royal Alberta Museum, Edmonton, AB, Canada}
\affil[3]{Department of Renewable Resources, University of Alberta, Edmonton, AB, Canada}
\affil[4]{Department of Mathematical and Statistical Sciences, University of Alberta, Edmonton, AB, Canada}
\affil[*]{Correspondence: pt1@ualberta.ca}
\date{}

\begin{document}

\maketitle

\newpage

\section{Abstract}

\begin{enumerate}
\item Spatial memory plays a role in the way animals perceive their environments, resulting in memory-informed movement patterns that are observable to ecologists. Developing mathematical techniques to understand how animals use memory in their environments allows for an increased understanding of animal cognition. 
\item Here we describe a model that accounts for the memory of seasonal or ephemeral qualities of an animal's environment. The model captures multiple behaviors at once by allowing for resource selection in the present time as well as long-distance navigations to previously visited locations within an animal's home range. 
\item We performed a set of analyses on simulated data to test our model, determining that it can provide informative results from as little as one year of discrete-time location data. We also show that the accuracy of model selection and parameter estimation increases with more location data. 
\item This model has potential to identify a specific mechanism in which animals use memory to optimize their foraging, by revisiting temporally and predictably variable resources at consistent time lags.
\end{enumerate}

\subsection{Keywords}

step selection function, cognitive map, spatial memory, grizzly bear, \textit{Ursus arctos}, animal movement, hidden Markov model

\section{Introduction}

Animal movement modelling has rapidly emerged as a subfield of ecology \citep{Nathan2008} due to advances in animal tracking \citep{Kays2015} and computational technology \citep{Kristensen2016}. The products of these advances have been widely applied to conservation and management \citep{Fortin2005, Graham2012, Gerber2019}. These models allow ecologists to understand the size and shape of an animal's home range \citep{Worton1989} as well as what habitat attributes animals prefer on a finer scale \citep{Gaillard2010}. To address the latter, ecologists have developed tools such as resource selection functions (RSFs; \citealp{Boyce1999}) and step selection functions (SSFs; \citealp{Fortin2005}). These allow for inference on an individual's habitat preference in what is known as third-order selection \citep{Johnson1980, Thurfjell2014}. The fine temporal and spatial resolution of these models allows ecologists to draw inference about a variety of behavioral processes, such as how an animal's movement rates are affected by its environment \citep{Avgar2016, Prokopenko2017} and how movement patterns change at different temporal scales \citep{OliveiraSantos2016, Richter2020}. And yet, even with the advances that have been made in animal movement modelling, some notable behavioral mechanisms are often not considered.

Spatial memory, defined by \cite{Fagan2013} as memory of the spatial configuration of one's environment, is one of the most important influences on animal movement patterns. The idea of episodic-like memory, which hypothesizes that animals can remember the ``what", ``where", and ``when" associated with specific events, is often intertwined within the intersection of spatial memory and foraging \citep{Munoz2009, Eacott2010, Allen2013, Crystal2018}. Many well-known behavioral processes, such as home range emergence \citep{VanMoorter2009, Riotte2015}, food caching \citep{Clayton1998}, and even migration \citep{Bracis2017, Merkle2019}, require the ability to remember the spatial location of landmarks or regions, which often requires some form of episodic-like memory of previous events. Animal species use spatial memory in different ways \citep{Fagan2013}, and the benefits an animal may receive from memory often depend on its environment \citep{Mueller2008, Mueller2011}. Theory on animal cognition has proposed that animals encode this spatial information in their brain as a cognitive map \citep{Tolman1948, OKeefe1978}. Ecologists have proposed multiple theories for the structure of these maps, with debate arising over whether a spatially explicit Euclidean map or a network-based topological map is more accurate \citep{Bennett1996, Sturz2006, Normand2009, Asensio2011}. The true structure of these cognitive maps in animals is still unclear and may vary in different animal species. In the least, a cognitive map is an effective mathematical vehicle to quantify how animals remember to revisit valuable places within their home ranges. The link between memory and movement has long interested ecologists (in the case of \citealp{Siniff1969}, for the purposes of home range modelling), but there are still ample opportunities for modelling.

Many animals experience seasonal variation within their home ranges \citep{Morey2007, Wiktander2001}, suggesting that a memory of these timings would be beneficial to optimize foraging. A key tenet of optimal foraging theory is that animals move to maximize their metabolic intake, and when the animal does decide to move, the timing of departure and the animal's subsequent destination are both important \citep{Charnov1976}. For example, primates time their journeys to previously visited resource patches with optimal feeding conditions as a means to maximize energetic intake \citep{Janmaat2006}. Sharks display intra-population variation and plasticity in their partially migratory movements, highlighting the viability of these long-distance navigations as an efficient foraging tactic \citep{Papastamatiou2013}. These recursive movement patterns are nearly impossible without some sort of seasonal episodic-like memory, where the animal must recall what it is foraging for, as well as when and where that food resource was last found \citep{Fagan2013}. Movement models that incorporate spatial memory can provide insight on how human-animal conflict \citep{Buderman2018}, habitat fragmentation \citep{Marchand2017}, and global warming \citep{Mauritzen2001} affect memory-informed animals.

Attempts to model these revisitations have proposed cognitive maps with spatial and temporal components, but have neglected to make inference about the specific nature of these influences. While many such approaches exist \citep{Dalziel2008, Avgar2013, Avgar2015, Vergara2016, Harel2018}, a common and simple technique involves integrating cognitive maps into SSFs \citep{Merkle2014, OliveiraSantos2016, Marchand2017}. A notable example is the model developed by \cite{Schlagel2014}, where cognitive maps are based on time since last visit (a form of episodic-like memory) for each point in space. It is proposed that animals will only be encouraged to revisit locations when they have not visited them recently, as seen in some ecological systems \citep{Davies1981}. This model was used to draw inference from gray wolf (\textit{Canis lupus}) movement patterns \citep{Schlagel2017}, but it does not provide detail on when animals choose to revisit portions of their home range. The model only considers the last visit to any point in space, disregarding any previous visits to that point. Time since last visit alone is insufficient to model the complex time-dependent spatial memory that inspires movement patterns described above, because waiting longer to revisit such locations may not always be beneficial for the animals (e.g., trees that lose their ripe fruit after too long).

Here we describe a model that mathematically estimates the timing and precision of these seasonally recursive movements (Fig. \ref{fig:schematic}). We employ innovative model fitting techniques \citep{Kristensen2016, Fischer2020} brought about by advances in computational methods to detect patterns in animal location data. Our modelling framework characterizes the movement of simulated or real animals according to four hypotheses: (N) the null hypothesis, assuming random walk behavior; (R) the resource-only hypothesis, assuming animals move entirely according to nearby resources without memory; (M) the memory-only hypothesis, assuming animals exhibit seasonal revisitation patterns within their home range with a prescribed mean lag time; and (RM) the resource-memory hypothesis, assuming animals are simultaneously influenced by local resources and spatial memory. This model expands on previous work, which has provided detail on how animals react to previously visited locations \citep{Schlagel2014}, how animals react to familiar locations at different times of day \citep{OliveiraSantos2016}, and how memory may decay over time \citep{Avgar2015}. \cite{Riotte2017} have even developed a movement metric capable of gauging how often recursions are present in animal movement data. We add to this rich array of literature by developing a model that quantifies how long animals may take to revisit certain regions of their home range, and how much a resource landscape plays a part in these movements, by analyzing the animal's entire movement path as opposed to the recursion events themselves. The model is not intended to answer the question of \textit{if} animals use memory, but instead \textit{how}, testing the prevalence of temporally consistent recursive movements in foraging animals.

To test our model, we first simulated movement tracks according to the model's prescribed rules on simulated environments, subsequently analyzing how sample size affects both model selection and parameter estimation. We found that even with data sizes equivalent to roughly one year of animal tracking data, the model accurately identified movement patterns consistent with the four different hypotheses and produced accurate parameter estimates. These results improved when tracks with more locations were simulated. We then fit the model to telemetry data from a population of Arctic grizzly bears (\textit{Ursus arctos}) and performed the same simulation analysis with real landscape data and movement parameters estimated for the bears. These bears live in a harsh environment where food resources are seasonal \citep{Edwards2015} and sparsely distributed \citep{Edwards2009}. We found a heavy influence of spatiotemporal memory in the bears' movement patterns, although we determined that more data may be required to analyze these populations than for simulated movements.

\section{Materials and Methods}

\begin{figure}
\centering
\includegraphics[width=\textwidth]{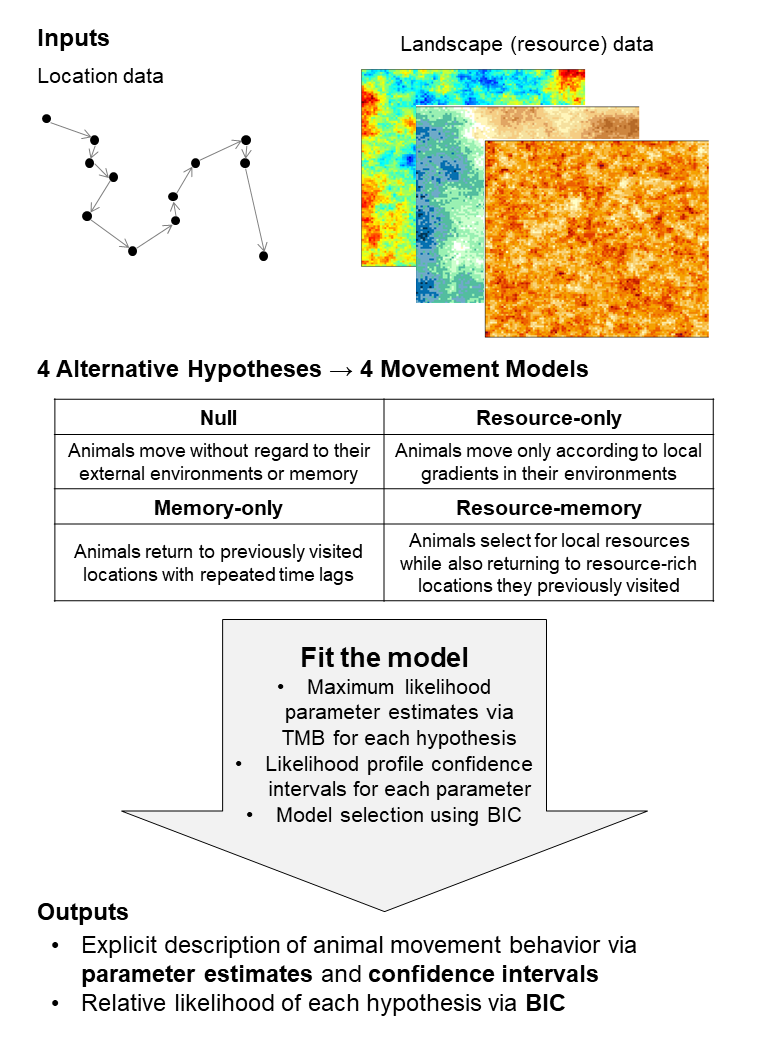}
\caption{Schematic describing our modelling framework. Given an animal's movement track, quantified as a set of spatial coordinates, as well as landscape data describing an animal's environment, we fit four nested, competing models using maximum likelihood estimation. The insight we gain from this process allowed us to make conclusions about the mechanistic drivers of animal behavior.}
\label{fig:schematic}
\end{figure}

Here we introduce a new modelling framework based on step selection functions that accounts for temporally consistent revisitations by animals that forage on ephemeral resources (Fig. \ref{fig:schematic}). We developed a nested structure of four models in discrete time and continuous space (see Table \ref{tab:pardescriptions} and Table S1 for a summary of the parameters and models) to address our four alternative hypotheses (N, R, M, RM). Our model fitting process, made possible through advanced automatic differentiation techniques, allows for further inference about the specific nature of these cognitive mechanisms. The novelty and complexity of the computational processes used to analyze animal location data with our model motivated multiple simulation-based studies to identify the statistical power and parameter estimability of our models.

\subsection{Modelling framework}

We fit a hidden Markov model (HMM) to animal movement data to incorporate switching between stationary (or quasi-stationary) and non-stationary states. HMMs are a first-order Markov process, implying that the animal's current state is entirely dependent on its most recent state. This approach is common in movement ecology due to the multitude of behavioral strategies observed in foraging animals \citep{Morales2004, Jonsen2013}. We employ this approach to differentiate resting or other stationary behavior from what the model would otherwise identify as spatial memory. Our model identifies time lags at which the animal moves particularly close to its previously visited locations, and staying put for one time step is interpreted mathematically by the model as strong recursive behavior with a time lag of one time step. Without including the stationary behavioral state, the model erroneously identifies this one-step time lag in most animal data.

An HMM consists of a Markov matrix $\mathbf{A}$ of state-switching probabilities as well as conditional probability distributions of the animal's spatial location for each state \citep{Jonsen2013}. For a model with $n$ different movement states, $\mathbf{A}$ maps from $\mathds{R}^n \rightarrow \mathds{R}^n$, with each column summing to 1. Our model has two states, so we can infer the structure of $\mathbf{A}$ from its diagonal. We denote these entries $\lambda$ and $\gamma$, representing the probability that the animal will stay in the stationary or non-stationary state, respectively, given it was just there. Explicitly, it takes the form below:

\begin{equation}\label{eq:A}
A = \begin{pmatrix}
\lambda & 1 - \lambda \\
1 - \gamma & \gamma
\end{pmatrix}
\end{equation}

While our model is meant to be applied to continuous-space animal data, we make an approximation by discretizing our landscape over a two-dimensional square grid. Empirical landscape data is rarely continuous in space, and the resolution of this data can suggest a clear choice for the resolution of the domain grid. We define points in continuous space as $\mathbf{x}$ (or $\mathbf{x}_t$ to represent the animal's location at time $t$) and their corresponding grid cells as $z$ or $z_t$. Thus, $\mathbf{x}_0 \in z_0$ is the animal's initial location.

We define our conditional probability density functions for the stationary and non-stationary state $f_s$ (which remains the same in all four models) and $f_{ns}$, respectively. Each conditional probability distribution represents a first-order Markov process modelling the animal's location $\mathbf{x}_t$ and its heading $\phi_t$ over time, which depend only on $\mathbf{x}_{t-1}$ and $\phi_{t-1}$ from the previous time step. Due to observation error in animal tracking data, we assumed that the animal's observed location may change slightly even if it is not moving \citep{Jonsen2013}, so we allowed for small ``movements" in our stationary state. The probability distribution for headings in the stationary state, $g_s(\phi_t| \phi_{t-1})$, is a uniform distribution since we assume no directional autocorrelation here, so

\begin{equation}\label{eq:gs}
g_s(\phi_t|\phi_{t-1}) = \frac{1}{2\pi},
\end{equation}

\begin{equation}\label{eq:condliks}
f_s(\mathbf{x}_{t}, \phi_t|\mathbf{x}_{t-1}, \phi_{t-1}, \rho_{s}) = \frac{2}{\pi \rho_s} g_s(\phi_t|\phi_{t-1}) \exp{ -\frac{\| \mathbf{x}_{t} - \mathbf{x}_{t-1}\|^2}{\pi \rho_s^2}}.
\end{equation}

We modeled the probability of the animal moving from $\mathbf{x}_{t-1}$ to $\mathbf{x}_t$ when in the stationary state using a half-Gaussian distribution with a fixed mean $\rho_s$. The half-Gaussian distribution has thinner tails than the more traditionally used exponential distribution, decreasing the probability of longer movements from this state. We fix $\rho_s$ to reduce model complexity, noting that it is fairly straightforward to do so based on the known degree of observation error or the resolution of environmental data.

\begin{figure}
\centering
\includegraphics[width=\textwidth]{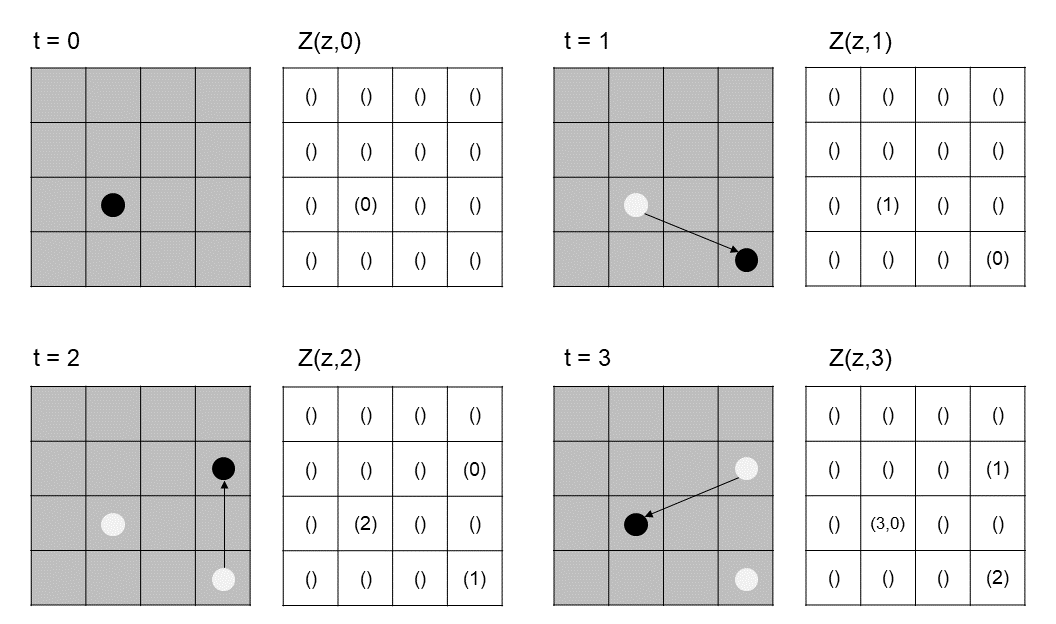}
\caption{Diagram describing how an animal's cognitive map $Z$ changes over four discrete time steps, given an animal's movement track, which is illustrated in the shaded panels. Each cell of $Z$ contains a linked list that starts out empty but is iteratively appended as the animal traverses its environment.}
\label{fig:cogmapdiagram}
\end{figure}

In the non-stationary state, we use a cognitive map structure to keep track of the animal's spatiotemporal movement experiences (Fig. \ref{fig:cogmapdiagram}). Our implementation of a cognitive map expands on the concept of time since last visit \citep{Davies1981, Schlagel2014, Schlagel2017} by allowing for the memory of more than just the last location to any point in space. Instead, we formulate the animal's cognitive map as the set of times since previous visits (TSPVs) for any area in space. This formulation allows for a form of seasonal episodic-like memory that expands on the ``time since last visit" framework \citep{Clayton1998, Martin2010}. We define this map $Z_t$ at each time $t$ as a function over the domain grid. At each grid cell $z$, $Z_t(z)$ is a linked list of integers, with each element of the list representing an animal's visit to a point inside that cell. $Z_0$ is a grid full of empty linked lists, except for $z_0$; $Z_0(z_0)$ is a list with one element, 0. We can obtain $Z_t$ if we know $Z_{t-1}$ as well as the animal's location at time $t$. When $t$ is incremented by 1, so is every entry on every linked list across the grid, and a new entry (0) is added to the linked list corresponding to the animal's new location:

\begin{equation}\label{eq:ziterative}
Z_t(z) = \begin{cases} 
Z_{t-1}(z) + 1 & \mathbf{x}_t \notin z \\
[Z_{t-1}(z) + 1, 0] & \mathbf{x}_t \in z
\end{cases}.
\end{equation}

\noindent where $[Z_{t-1}(z) + 1, 0]$ implies adding 1 to every entry of the linked list $Z_{t-1}(z)$ and appending it with a new value 0.

The function $f_{ns}$, which models the animal's location and heading in the non-stationary state, resembles a step selection function \citep{Fortin2005, Forester2009}, with two main components: $k$, the resource-independent movement kernel; and $W$, the environmental (or cognitive) weighting function. The function $k$ describes the animal's locomotive capability while $W$, which may depend on the animal's cognitive map $Z_{t-1}$, describes how attractive the point is to the animal. This yields the following expression for $f_{ns}$:

\begin{align}\label{eq:condlikns}
f_{ns}(\mathbf{x}_{t},\phi_t & |\mathbf{x}_{t-1}, \phi_{t-1}, Z_{t-1}, \Theta_{1}, \Theta_{2}) \nonumber \\
& = \frac{k(\mathbf{x}_{t}|\mathbf{x}_{t-1}, \phi_{t-1}, \Theta_{1})W(\mathbf{x}_{t}|Z_{t-1}, \Theta_{2})}{\int_\Omega k(\mathbf{x'}|\mathbf{x}_{t-1}, \phi_{t-1}, \Theta_{1})W(\mathbf{x'}|Z_{t-1}, \Theta_{2})d\mathbf{x'} }.
\end{align}

\noindent The integral in the denominator serves as a normalization constant to ensure that $f_{ns}$ integrates to 1. The parameter vector $\Theta_2$ represents parameters related to the $W$ and $\Theta_1$ represents the locomotive parameters associated with $k$, namely $\rho_{ns}$ which describes the animal's mean step length and $\kappa$ which describes the degree of directional autocorrelation in the animal's movements. For each of our four models (null, resource-only, memory-only, resource-memory), the animal's resource-independent movement kernel $k$ (as well as $\Theta_1$) has the same formulation. We modeled the distance between $\mathbf{x}_t$ and $\mathbf{x}_{t-1}$, known as a step length, using an exponential distribution with mean parameter $\rho_{ns}$, and modeled the heading $\phi_t$ using a von Mises distribution centered at $\phi_{t-1}$ with concentration parameter $\kappa \geq 0$ (Equation \ref{eq:gns}). Higher values of $\kappa$ indicate straighter movement. We assume here that the animal's step lengths and turning angles are independent. This modelling structure, known more generally as a correlated random walk, has been applied to a variety of ecological systems \citep{Fortin2005, Auger2015, Duchesne2015}, and the exponential and von Mises distributions are both particularly easy to deal with analytically while still providing accurate fits for a majority of field data \citep{Codling2008, Thurfjell2014}. We formulate $k$ such that

\begin{equation}\label{eq:gns}
g_{ns}(\phi_t|\phi_{t-1}) = \frac{\exp{(\kappa \cos(\phi_t - \phi_{t-1}))}}{2 \pi I_0(\kappa)}, \textrm{ and}
\end{equation}

\begin{equation}\label{eq:kgeneral}
k(\mathbf{x}_{t}|\mathbf{x}_{t-1}, \phi_{t-1}, \Theta_{1}) = \frac{\exp{\left( -\frac{\|\mathbf{x}_{t} - \mathbf{x}_{t-1}\|}{\rho_{ns}} \right)}}{\rho_{ns}} g_{ns}(\phi_t|\phi_{t-1}),
\end{equation}

\noindent where $I_0(\kappa)$ is the modified Bessel function of order 0. Notice that $\phi_t$, the animal's heading at time $t$, is not explicitly included in the left side of Equation \ref{eq:kgeneral}; it can instead be calculated easily if $\mathbf{x}_t$ and $\mathbf{x}_{t-1}$ are known \citep{Fortin2005}. Note that $g_{ns}$, just like $g_s$, is separate from the rest of $k$, following our assumption that the animal's step lengths and bearings are independent.

The only mathematical difference between the four models is the formulation of $W$. To differentiate between these different formulations, we refer to them as $W_N$, $W_R$, $W_M$, and $W_{RM}$ for the null, resource-only, memory-only, and resource-memory models, respectively. The set of parameters we estimate in each model also varies, so we define $\Theta_{2, N}$, $\Theta_{2, R}$, $\Theta_{2, M}$, and $\Theta_{2, RM}$ in a similar respect.

\subsubsection{Null model}

The null model describes an animal's locomotive capability and directional autocorrelation based on its observed movement track. As a result, there is no extra weighting, so $W_N(\mathbf{x}_{t}|\Theta_{2, N}) = 1$ for all $\mathbf{x}_t$ in space, and $\Theta_{2, N}$ is the empty set. As a result, when considering the null model, $f_{ns}$ is equal to $k$.

\subsubsection{Resource-only model}

The resource-only model has the following key component:

\begin{itemize}
\item[(R1)] the animal's movement is driven by third-order selection for resources reachable within one time step. 
\end{itemize}

\noindent As a result, $W_R$ resembles the weighting function from an RSF or SSF \citep{Boyce1999, Fortin2005}. If we are interested in $P$ different resource covariates (expressed mathematically at each spatial location $\mathbf{x}$ as $r_1(\mathbf{x}), ..., r_P(\mathbf{x})$), we must estimate selection parameters $\beta_1, ..., \beta_P$ for each covariate. These parameters make up $\Theta_{2, R}$. The expression for our weighting function in the resource-only model is a linear combination of the covariates:

\begin{equation}\label{eq:wres}
W_R(\mathbf{x}_{t}|\Theta_{2, R}) = \exp\Bigg[\sum_{p=1}^P \beta_p r_p(\mathbf{x}_t)\Bigg].
\end{equation}

\subsubsection{Memory-only model}

The memory-only model contains the following key components: 

\begin{itemize}
\item[(M1)] the animal uses a cognitive map to remember the timing of previous visits to regions of its environment, and
\item[(M2)] it will return to locations it previously visited after temporally similar time lags.
\end{itemize}

\noindent This type of cognitive map has been supported in the literature \citep{Normand2009, Martin2010, Schlagel2014} as has the validity of path recursions and revisitations as a foraging strategy for animals \citep{Berger2015, Schlagel2017}. Note that this behavior could arise from multiple mechanisms: if an animal is foraging for periodically available resources, we can use its previous locations to determine where it might be in the future, and if an animal forages on some depleting resource, we could use this model to identify how long the animal waits before returning to a resource it had previously depleted. Note, though, that the memory-only model assumes a homogeneous landscape, as resource data are not included. While this assumption is usually unrealistic, we include it as an alternate hypothesis to models including resource selection. In cases where appropriate resource data are not available, or the existing resource data are insufficient to explain patterns in the movement data, the memory-only model serves to identify if a pattern of timed re-visitation exists.

We calculate $W_M$ based on distance to previously visited points on the animal's track. Given some time lag $\tau$, we can use the cognitive map $Z_t$ to find the point in space (or at least, the grid cell) where the animal was $\tau$ time indices ago. There is always exactly one grid cell $z_{t - \tau}$ where $\tau$ is an element of the linked list $Z_t(z_{t - \tau})$.

For each time lag $\tau$, we compute the distance between the animal's current location and $z_{t - \tau}$, $\| \mathbf{x}_t - z_{t - \tau}\|$, and transform it using an exponential decay function with decay parameter $10^{\alpha}$. The primary role of $\alpha$ is to convert distances to unitless quantities representing attractiveness. Under the assumption that points closer to previously visited locations are more attractive, we use $\exp{(-10^{\alpha} \|\mathbf{x}_t - z_{t - \tau}\|)}$ as the transformation for the distance between $\mathbf{x}$ and the center of $z_{t - \tau}$. We include the power here so $\alpha$ can be any real number, and use 10 so its estimate can be interpreted more easily. The transformation with $\alpha$ produces a discounting of importance with distance, where $\alpha$ quantifies how quickly this importance is discounted spatially. If $\alpha$ is larger, then points must be very close to the previously visited location for the animal to deem them attractive. As $\alpha$ decreases, the mathematical difference between a step 1000 m away and a step 2000 m away is amplified, suggesting that the animal understands these differences in space on a wider scale. The value of $\alpha$ may be informative about the heterogeneity of the landscape, which can be informative about how animals value the importance of distance in predicting resource quality \citep{Farnsworth1999}.

The animal's revisitation schedule, which is mediated by two parameters $\mu$ and $\sigma$, dictates the weights for each of these exponentially transformed distances. The timing with which an animal navigates back to an existing location can be thought of as a random process, following a Gaussian distribution with mean parameter $\mu$ and standard deviation parameter $\sigma$. We can imagine that this timing reflects the state of the environment, with $\mu$ indicating the time scale at which resources may come and go and $\sigma$ indicating the variability of these revisitations. For any given time lag $\tau$, the exponentially transformed distance between $\mathbf{x}_t$ and $z_{t-\tau}$ is weighted by the Gaussian probability distribution function $\varphi(\tau|\mu, \sigma)$. This produces a weighted mean of exponentially transformed distances, following the hypothesis that animals will navigate towards points they visited roughly $\mu$ time increments ago; the most ``attractive" points for the animal are closest to $z_{\mu}$. We introduce one final parameter, $\beta_d$, a ``selection coefficient" for memorized locations. This parameter can be thought of as the relative probability of revisiting a memorized location instead of moving randomly or selecting for present-time resources. We restricted $\beta_d \geq 0.5$ (implying $\log \frac{\beta_d}{1 - \beta_d} > 0$), in line with the hypothesis that animals select for (not against) previously visited locations.

The resulting formulation of $W_M$ is as follows:

\begin{align}\label{eq:wmem}
W_M(\mathbf{x}_{t}|& Z_{t-1}, \Theta_{2, M}) \nonumber \\ & = \exp\Bigg(\tilde{\beta}_d \Bigg[\frac{\sum_{\tau=1}^{t}\varphi(\tau|\mu, \sigma)\exp{(-10^{\alpha} \|\mathbf{x}_{t} - z_{t-\tau}\|)}}{\sum_{\tau=1}^{t}\varphi(\tau|\mu, \sigma)}\Bigg]\Bigg),
\end{align}

\noindent where $\tilde{\beta}_d = \log(\frac{\beta_d}{1 - \beta_d})$, and $\Theta_{2,M}$ contains $\mu, \sigma, \beta_d$, and $\alpha$.

$W_M$ does not directly contain any periodic components (the Gaussian weight simply just has one mode around $\mu$), and we do this to increase the flexibility of the model. In the event that an animal is pursuing resources that vary periodically with a period of $\mu$, its location at any point is likely to be nearby its location $\mu$ timesteps ago. Movements simulated from this model do also produce movements that are somewhat periodic, although the spatial correlation between an animal's location and its location $\mu$ time steps prior is stronger than locations separated temporally by $2\mu$, for example.

\subsubsection{Resource-memory model}

The resource-memory model incorporates both resource selection and memory into the animal's movements, so (R1) and (M1) still remain as components in this model. However, there is one additional component that is not present in the resource-only or memory-only models:

\begin{itemize}
\item[(RM1)] the animal will return to locations it previously visited at a prescribed and scheduled time if habitat conditions there were favorable; otherwise it will avoid these areas.
\end{itemize}

\noindent Models combining resources and memory in some way have proven to be effective in explaining movement patterns for many different animals \citep{Dalziel2008, Merkle2014, Schlagel2017}. The resource-memory model builds on the memory-only model, which is often unrealistic due to the omission of environmental data, by truly quantifying an animal's episodic-like memory, capturing the ``when" and ``where" of an animal's spatial experience via $Z$ and augmenting this with the ``what": the resource quality at these previously visited points. The addition of hypothesis RM1 produces memory that is resource-dependent, whereas the memory-only model works under the typically false simplifying assumption of a spatially homogeneous landscape.

The linear combination of resource covariates $\sum_{p=1}^{P}\beta_p r_p(\mathbf{x})$ is relative, so we introduced an additional parameter $\beta_0$ representing the relative probability of visiting a faraway location depending on its resource quality. As $\beta_0$ approaches 1, the animal perceives all previously visited locations as ``attractive" for revisitation. We transform this parameter with an inverse logistic function so it represents a pseudo-intercept (recall that traditional SSFs and are conditional models and do not require an intercept; \citealp{Fortin2005}).

The weighting function now includes present-time resource selection in the first sum and memorized information in the second term:

\begin{align}\label{eq:wcomb}
& W_{RM}(\mathbf{x}_{t}|Z_{t-1}, \Theta_{2, RM}) = \exp \Bigg( \sum_{p=1}^{p} \beta_p r_p(\mathbf{x}_{t}) + \\ \nonumber &  \tilde{\beta}_d\Bigg[\frac{\sum_{\tau=1}^{t}\varphi(\tau|\mu, \sigma) \exp{(-10^{\alpha} \|\mathbf{x}_{t} - z_{t-\tau}\|)}(\tilde{\beta}_0 + \sum_{p=1}^{P}\beta_p r_p(z_{t-\tau}))}{\sum_{\tau=1}^{t}\varphi(\tau|\mu, \sigma)} \Bigg] \Bigg),
\end{align}

\noindent where $\tilde{\beta}_d = \log(\frac{\beta_d}{1 - \beta_d})$ and $\tilde{\beta}_0 = \log(\frac{\beta_0}{1 - \beta_0})$.

\begin{table}[h]
\centering
{\renewcommand{\arraystretch}{1.2}
\begin{tabular}{|c|c|c|c|c|c|c|}
\hline
& \textbf{Units} & \textbf{Description} & \textbf{N} & \textbf{R} & \textbf{M} & \textbf{RM} \\
\hline
$\rho_{ns}$ & $\frac{\textrm{distance}}{\textrm{time}}$ & Mean movement speed in non-stationary state & X & X & X & X \\
\hline
$\kappa$ & N/A & Degree of directional autocorrelation & X & X & X & X \\
\hline
$\beta_0$ & N/A & Probability of revisitation & & & & X \\
\hline
$\beta_i$ & $\frac{1}{r_i \textrm{ units}}$ & Resource selection coefficient(s) & & X & & X \\
\hline
$\beta_d$ & N/A & Strength of selection for memorized areas & & & X & X \\
\hline
$\mu$ & time & Mean time lag between revisitations & & & X & X \\
\hline
$\sigma$ & time & Standard deviation in time between revisitations & & & X & X \\
\hline
$\alpha$ & $\log(distance)$ & Degree of perceptual resolution & & & X & X \\
\hline
$\lambda$ & N/A & Probability of staying in stationary state & X & X & X & X \\
\hline
$\gamma$ & N/A & Probability of staying in non-stationary state & X & X & X & X \\
\hline
\end{tabular}}
\caption{Description of model parameters, including units (N/A implies that the parameter is unitless) and models (N = null; R = resource-only; M = memory-only; RM = resource-memory) in which the parameters are estimated. For functions and other quantities that were not fit as model parameters, see Table S1.}
\label{tab:pardescriptions}
\end{table}

The null model is a special case of both the resource-only and memory-only models, which are both a special case of the resource-memory model. Setting $\beta_i = 0$ for $i = 1, 2, ..., P$ and $\log\left(\frac{\beta_0}{1 - \beta_0}\right) = 1$ in the resource-memory model yields the memory-only model, while setting $\beta_d = 0$ yields the resource-only model. Nesting models is advantageous for many mathematical reasons, including the ability to conduct likelihood ratio tests between models \citep{Burnham2004}.

\subsection{Statistical inference}

We fit the four models to discrete-time, continuous-space animal movement data and used information theory to identify which corresponding hypothesis was most likely to be true. We identified the optimal set of parameters for a given track using maximum likelihood estimation, and used likelihood profiling to obtain accurate confidence intervals for our parameters.

\subsubsection{Likelihood function}

The likelihood of a set of model parameters for one step is a weighted sum of the conditional likelihood functions ($f_s$ and $f_{ns}$), weighted by the probability of being in each state. These state probabilities depend on probabilities for the previous step, so for the first point we fit (there is no previous step), we fixed $\delta_s$, the probability of being in the stationary state right before the data begins, as the proportion of steps shorter than $\rho_s$.

The likelihood function for the entire track is a product of the likelihoods for each step included in model fitting. We omitted all animal locations before some time $t^*$, since our model (or at least, the memory-only and resource-memory models) relies on past information to explain where the animal may go. We left the portion of the track that happened before $t = t^*$ to ``train" the model on what the animal remembers. Thus, our iterative formula for the likelihood function begins at $t = t^*$. We define $\mathbf{\Phi}_t \in \mathds{R}^2$ as the vector of state probabilities for time $t \geq t^*$, and we calculate our likelihood using the iterative equations below: 

\begin{align}\label{eq:phi}
\mathbf{\Phi}_{t^*} &=  (\delta_s, 1-\delta_s)^T, \\
\mathbf{P}_t &= \begin{pmatrix}f_s(\mathbf{x}_{t}|\mathbf{x}_{t-1}, \rho_s) & 0 \\ 0 & f_{ns}(\mathbf{x}_{t}, \phi_t|\mathbf{x}_{t-1}, \phi_{t-1}, Z_{t-1}, \Theta_{1}, \Theta_{2})\end{pmatrix}, \\
\mathbf{\Phi}_t &= \frac{\mathbf{\Phi}_{t-1}^T \mathbf{P}_{t-1}}{\|\mathbf{P}_{t-1} \mathbf{\Phi}_{t-1}\|} \mathbf{A}.
\end{align}

\noindent Then, following \cite{Whoriskey2017}, the overall likelihood for the model is $\prod_{t=t^*}^{t_max} \mathbf{\Phi}_t^T \mathbf{P}_t \mathbf{1}$, where $\mathbf{1} = (1, 1)^T$.

We approximate the denominator of Equation \ref{eq:condlikns} with a sum so we do not have to integrate every time we evaluate the likelihood function. As is commonly done with SSFs \citep{Thurfjell2014}, we calculated $W$ at a set of ``control points" for each observed point $\mathbf{x}_t$. If $\mathbf{x}_t$, the endpoint of a step from $\mathbf{x}_{t-1}$, is a random variable conditional on $Z_{t-1}$, $\Theta_1$, and $\Theta_2$, the integral in the denominator of Equation \ref{eq:condlikns} is $E(W(\mathbf{x}_t))$. Thus, we can approximate it by estimating the mean value of $W$ at a set of simulated draws from $\mathbf{x}_t$, which has probability density function $k$. This gives us the following approximation for $f_{ns}$:

\begin{align}\label{eq:ssfapprox}
\tilde{f}_{ns}(\mathbf{x}_{t},\phi_t & |\mathbf{x}_{t-1}, \phi_{t-1}, Z_{t-1}, \Theta_{1}, \Theta_{2}) \nonumber \\
& = \frac{k(\mathbf{x}_{t}|\mathbf{x}_{t-1}, \phi_{t-1}, \Theta_{1})W(\mathbf{x}_{t}|Z_{t-1}, \Theta_{2})}{\frac{1}{K}\sum_{j=1}^{K} W(\mathbf{x}_{t,j}|Z_{t-1}, \Theta_{2})},
\end{align}

\noindent where $\mathbf{x}_{t,j}$ represents the $j^{th}$ control point (a simulated step starting at $\mathbf{x}_{t-1}$) and $K$ is the number of control points per observed step. From this approximation, it becomes evident that each model compares $W$ from steps the animal actually took to steps that are simulated from a random walk. This implies that if an animal occasionally returns to previously visited locations as a result of random movement, the model will account for this and identify the null model as a more parsimonious explanation of the data than the other models. For the memory-only model to truly be an effective explanation of movement patterns observed in the data, these revisitations must be frequent and temporally consistent.

\subsubsection{Fitting the model}

We fit the model to data using maximum likelihood estimation, with the Template Model Builder (TMB) R package \citep{Kristensen2016} improving numerical accuracy for this complex problem. TMB has been used to fit complex animal movement models, including HMMs \citep{Albertsen2015, Auger2017, Whoriskey2017}. TMB uses automatic differentiation to calculate the gradient of a multidimensional likelihood function using pseudo-analytical methods, as opposed to traditional finite-difference methods that are slow and frequently result in numerical errors \citep{Skaug2006}. We wrote a likelihood function for each model in C++, which TMB compiles and returns as a callable function in R \citep{Kristensen2016}. This allowed us to use an R optimizer of our choice while also benefiting from C++'s superior programming speed.

We used the R \textit{nlminb} function to obtain maximum likelihood estimates for the negative log of our likelihood function. To prevent our model from producing errors or unrealistic results, we imposed various bounds on some of the parameters. We bounded the estimation for $\mu$ at $t^*$ because if $\mu > t^*$, we would not be able to identify a signal due to a lack of training data. We also put a lower bound on $\sigma$; when this parameter was small, the partial derivative of our likelihood function with respect to $\mu$ became noisy, leading to computational errors in optimization. We found that a lower bound of approximately 20 time indices eliminated this problem. We additionally required estimates for $\alpha < -\log_{10}(\bar{\rho})$, where $\bar{\rho}$ is the animal's empirical mean step length (for context, we expect $\bar{\rho}$ to be close to but slightly smaller than $\rho_{ns}$). Values of $\alpha$ above this bound imply that the animal cannot perceive a difference between a few step lengths, which is unreasonable biologically. For parameters with fairly restrictive bounds ($\lambda$, $\gamma$, $\beta_d$, and $\beta_0$, which are bounded between 0 and 1), we performed logit transformations ($\tilde{\lambda} = \log\frac{1}{1-\lambda}$, for example) so the optimizer would more effectively traverse the parameter space.

We tested two ``initial values" for $\mu$ for each dataset we fit the model to, picking the fit that gave us the best likelihood function value. When profiling the likelihood surface with respect to this parameter, we often found many local optima, so we fit the model with initial values of $t^*$ and $\frac{t^*}{2}$. Fitting with different initial values incurs additional computational time (we are effectively running the optimization algorithm twice) but is necessary due to the importance of picking a good initial value for each parameter \citep{Pan1998}. Using a different number of initial values for $\mu$ may be advantageous for some datasets.

For a model as complicated as this one, obtaining confidence intervals (CIs) using traditional Wald-type methods does not always produce accurate results. We frequently found this to be true for our model in practice so we used the likelihood profiling from \cite{Fischer2020}. Given a multidimensional objective function with a known optimum, this algorithm finds confidence intervals for one parameter at a time by performing a binary search algorithm for a target function value (typically, the optimum minus some small confidence threshold). The algorithm starts searching at the optimal parameter value, and tries an initial step, fixing the parameter in question at this value and optimizing the rest of the function parameters. This process is repeated subsequently until the lengths of each step in parameter space are small enough for the algorithm to converge \citep{Fischer2020}.

We used the Bayesian Information Criterion (BIC) to rank the four models by their likelihood and identify the hypothesis that was most likely to be true. BIC has a stronger penalization for model complexity than the more commonly used Akaike Information Criterion (AIC), and is a more useful criterion for model selection when one is interested in the truth of a hypothesis rather than the predictability of a model \citep{Burnham2004}.

\subsection{Simulation studies}

Before applying our model to an ecological system, we simulated data and used it to test the model. These simulations are individual-based representations of our model that produce movement patterns associated with our four hypotheses. We performed this analysis as a means to ensure that our fitting methods could accurately identify the parameter values prescribed by the model. At each time index, we used our Markov matrix $\mathbf{A}$ to decide whether the animal would change its behavioral state. If the animal was in the stationary state we simulated a random step from $f_s$ (half-Gaussian step length, uniform turning angle). For the non-stationary state, we simulated from $f_{ns}$ using a Monte Carlo method \citep{Parzen1961}. We first calculated $W$ for the entire grid, then we simulated a large number of random steps from $k$ (Equations \ref{eq:gns} and \ref{eq:kgeneral}). This simulation process resembles the generation of control points in Equation \ref{eq:ssfapprox}, but we simulated $N_r = 10000$ steps at each point in time. Making $N_r$ very large did not greatly affect computational time, so we did so in the interest of accurately approximating Equation \ref{eq:condlikns}. These simulations took place on a bounded grid representing the hypothetical landscape, and any of the $N_r$ proposed steps that took the animal off this grid were re-sampled until they were on the grid. While this resembles reflective boundary conditions, the animal is not assumed to ``bounce off" the boundary or interact with it in any way other than avoiding it. Note that it is possible to tune the animal's mean step length as well as the size of the landscape in simulations to drastically reduce the probability of this happening, which we did. We then randomly choose one of the steps based on the values of $W$ at each step, with the probability of any step $\mathbf{x}_{t,i}$ being chosen described below:

\begin{equation}\label{eq:stepchoice}
\frac{W(\mathbf{x}_{t,i}|Z_{t-1}, \Theta_{2})}{\sum_{j=1}^{N_r} W(\mathbf{x}_{t,j}|Z_{t-1}, \Theta_{2})}.
\end{equation}

\noindent For models that incorporate memory, we simulated memoryless training data ($W_M = W_N$ for the memory-only model, and $W_{RM} = W_R$ for the resource-memory model) for $t < t^*$. As expected, these initial points are omitted from model fitting.

\subsubsection{Model verification: simulated data}

We simulated tracks on artificial landscapes with preset model parameters, then fit the model to these tracks to explore parameter estimability and model selection accuracy. We varied the length of these tracks, $T = t_{max} - t^*$, as well as $K$, the number of controls points per step, to evaluate the amount of data required for accurate inference. Specifically, we tested four ``treatment groups":  $T = 600, K = 10; T = 600, K = 50; T = 1200, K = 10;$ and $T = 1200, K = 50$. 

We used the R NLMR package \citep{Sciaini2018} to simulate spatially autocorrelated Gaussian random fields representing our resource covariates. For each treatment group, we simulated 50 random movement tracks for each hypothesis. Each group of 50 tracks had the same set of parameters. In our simulations, we simulated environments for $P = 3$ resource covariates per track using the \textit{nlm\_gaussianfield} function in R. We then fit all four models to each track individually, then used BIC to identify how often the ``correct" model was selected for each movement track. We compared these results with AIC to confirm that BIC is the most suitabile information criterion for our modelling framework. We also estimated the bias and mean squared error (MSE; the mean squared difference between the parameter estimate and the true value) for each parameter with each model.

\subsubsection{Model application: grizzly bear case study}

We applied the model to grizzly bears in the Canadian Arctic, and then repeated the simulation study with data and model parameters from this system. Bears were captured from 2003 to 2006 and released with global positioning system (GPS) collars. Collars returned a location every four hours while the bear was not hibernating, and remained on the bears for up to four years \citep{Edwards2009}. The University of Alberta Animal Care and Use Committee for Biosciences approved all animal capture and handling procedures, which were in accordance with the Canadian Council on Animal Care. Capture and tracking was conducting under permit from the Government of the Northwest Territories, Department of Environment and Natural Resources, Inuvik Office (Permit numbers: WL3104, WL3122, WL3282, WL5352, and WL5375) following methods approved by the University of Alberta Animal Care and Use Committee for Biosciences (Permit numbers: ACUC412305, ACUC412405, ACUC412505, ACUC412605, and ACUC412705) in accordance with the Canadian Council on Animal Care guidelines.

The bears were collared in the Mackenzie River Delta region in the Northwest Territories \citep{Edwards2009}. Resources in the region are sparse and heterogeneous both in space and time \citep{Shevtsova1995, Edwards2015}. To survive and forage optimally, bears take advantage of ephemeral, unpredictable, or seasonally available resources through a variety of foraging strategies \citep{Edwards2009, Edwards2011, Edwards2015}.

We analyzed grizzly bear habitat selection using multiple sources of environmental data describing the Mackenzie Delta region. Vegetation class data for the region assigned a one of 46 classes (indicating the dominant plant type or terrain) to each 30x30 m cell. A digital elevation model for the region (with 30x30 m cell resolution) provided information on elevation and slope. We also used an RSF layer estimating resource use for Arctic ground squirrels (\textit{Urocitellus parryii}), a common grizzly bear prey item \citep{Barker2010, Edwards2015}. We considered $P = 6$ resource covariates: berry density, represented as a likelihood of having berries for each vegetation class; distance to turbid water, an indicator of broad whitefish (\textit{Coregonus nasus}; a grizzly bear prey item; \citealp{Barker2009}) density as well as riparian habitat; Arctic ground squirrel density, taken directly from the RSF; sweetvetch (\textit{Hedysarum alpinum}; a key grizzly bear food item; \citealp{Edwards2015}) density, estimated by the vegetation class data; distance to the nearest of two towns in the region; and distance to six remote industrial camps (likely with little human activity). We modelled these resources in two different ways, fitting the resource-only and resource-memory models twice to each bear with different interpretations: resources constant in time and resources that explicitly vary throughout the year. We expected that if the movement patterns we had observed were simply a result of the resource variation, as opposed to the bears memorizing the location and timing of these resources, then the resource-only model with seasonal resources would outperform any of our models including memory. We defined an interval of availability based on the literature \citep{Macdonald1995, Buck1999, Gau2002, MacHutchon2003}, and assigned the value 0 to every point on the grid outside the time interval for that resource. The null and memory-only model, which do not incorporate resources, are unaffected by this change, but since we needed to generate new available points for the seasonally varying resources, the model fits were slightly different for these models as well.

Of the 21 bears with enough data for model fitting (at least two years of GPS collar data), we selected the eight with the most GPS fixes (these bears had at least three years of collar data). We set $\rho_s = 30$ meters, corresponding to the length of one grid cell for the environmental raster data, and we set $t^* = 365$ days. We used $K = 50$ control points when fitting the models. For each of these bears, we fit the models to the entire track as well as each year individually, comparing model selection between years. We then replicated that analysis using simulated bear tracks; for each bear, we simulated 100 movement tracks using the optimal parameters for each bear and the Mackenzie Delta environmental data. We simulated tracks of length $T = 600$ (approximately one year of grizzly bear GPS data, accounting for missed fixes and hibernation) and $T = 1200$ to evaluate how model selection accuracy changed with sample size. We used BIC to identify the hypothesis that most accurately explained each movement track, and also conducted likelihood ratio tests for each pair of nested models to determine the significance of specific behavioral signals.

\section{Results}

Our modelling structure allows ecologists to explain movement patterns identified from location data according to a set of four hypotheses, of which two incorporate complex time-dependent spatial memory. For animals that appear to use memory, our parametric approach evaluates the temporal consistency of navigations to previously visited locations in an animal's home range. By fitting the model to simulated data we showed that the accuracy of the model is improved by sample size, and ecologists can also increase parameter estimability by simulating additional control points. Still, the amount of data required to draw accurate inference from the model is not large, as we show both with simulated environments and real-life landscape data (where the model is slightly less accurate).

\subsection{Model verification: simulated data}

The model's ability to accurately characterize each type of movement behavior increased with the amount of location data ($T$) but not with the number of control points ($K$; Table \ref{tab:biccountssims}). The model identified null and resource-only movements accurately at all treatment levels, but the model's ability to identify memory-only and resource-memory movement increased for longer simulated tracks. As a whole, increasing $K$ does not improve model selection accuracy for either choice of $T$. The most common misidentification at all sample sizes was mistaking resource-memory movement for resource-only or memory-only movement. 

\begin{table}[h!]
\centering
\begin{tabular}{|c|c||c|c|c|c||c|c|c|c|}
\hline
\multicolumn{2}{|c|}{\multirow{2}{*}{}} & \multicolumn{4}{|c|}{$K = 10$} & \multicolumn{4}{|c|}{$K = 50$} \\
\cline{3-10}
\multicolumn{2}{|c|}{} & N & R & M & RM & N & R & M & RM \\
\hline
\hline
\multirow{4}{*}{\rotatebox[origin=c]{90}{$T = 600$}} & N & \cellcolor{lightgray} 48 & \cellcolor{lightgray} 0 & \cellcolor{lightgray} 0 & \cellcolor{lightgray} 2 & 47 & 0 & 0 & 3 \\
\cline{2-10}
& R & \cellcolor{lightgray} 0 & \cellcolor{lightgray} 45 & \cellcolor{lightgray} 0 & \cellcolor{lightgray} 5 & 0 & 46 & 0 & 4 \\
\cline{2-10}
& M & \cellcolor{lightgray} 7 & \cellcolor{lightgray} 0 & \cellcolor{lightgray} 40 & \cellcolor{lightgray} 3 & 4 & 0 & 44 & 2 \\
\cline{2-10}
& RM & \cellcolor{lightgray} 0 & \cellcolor{lightgray} 5 & \cellcolor{lightgray} 7 & \cellcolor{lightgray} 38 & 0 & 8 & 7 & 35 \\
\hline
\hline
\multirow{4}{*}{\rotatebox[origin=c]{90}{$T = 1200$}} & N & 49 & 0 & 0 & 1 & \cellcolor{lightgray} 45 & \cellcolor{lightgray} 0 & \cellcolor{lightgray} 0 & \cellcolor{lightgray} 5 \\
\cline{2-10}
& R & 0 & 46 & 0 & 4 & \cellcolor{lightgray} 0 & \cellcolor{lightgray} 47 & \cellcolor{lightgray} 0 & \cellcolor{lightgray} 3 \\
\cline{2-10}
& M & 2 & 0 & 47 & 1 & \cellcolor{lightgray} 0 & \cellcolor{lightgray} 0 & \cellcolor{lightgray} 50 & \cellcolor{lightgray} 0 \\
\cline{2-10}
& RM & 0 & 3 & 2 & 45 & \cellcolor{lightgray} 0 & \cellcolor{lightgray} 7 & \cellcolor{lightgray} 2 & \cellcolor{lightgray} 41 \\
\hline
\end{tabular}
\caption{Breakdown of model selection counts using BIC for the simulated tracks. The row represents the ``true" model that the tracks were simulated from (N = null; R = resource-only; M = memory-only; RM = resource-memory), while the column represents the model that was identified as the most parsimonious explanation of the data using BIC. Treatment groups (based on T, the length of the fitted movement track, and K, the number of available points per timestep)} are identified by the outer left and upper portions of the table and are separated by shading.
\label{tab:biccountssims}
\end{table}

Using AIC instead of BIC resulted in a higher rate of ``false positives" for memory (i.e., the resource-memory or the memory-only model was identified as the most parsimonious explanation for memoryless simulated tracks), and made model selection less accurate overall (Appendix: Table S2). Likelihood ratio tests on the same dataset for each pair of nested models revealed a similar trend; the likelihood ratio test often identified memory when it was not incorporated into the simulated tracks (Appendix: Table S3).

\begin{figure}
    \centering
    \includegraphics[width=\textwidth]{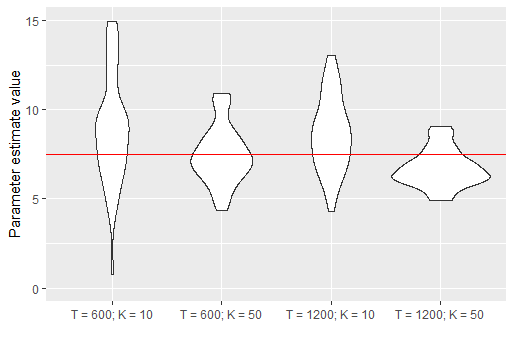}
    \caption{Violin plot of parameter estimates for $\beta_1$ in the resource-memory model for our four treatment groups (listed on the x-axis), with 50 simulations per plot. The true value of 7.5 is denoted by a horizontal red line.}
    \label{fig:beta1violin}
\end{figure}

The model produced more accurate parameter estimates with larger values of $T$ and $K$ (Table \ref{tab:biasmsecomb}). When focusing on $\beta_1$ in the resource-memory model, we can see that bias does not change as much with different treatment groups as MSE (Fig. \ref{fig:beta1violin}). For the simpler movement parameters ($\rho_{ns}$, $\kappa$, $\lambda$, $\gamma$), parameter estimates were consistent even with smaller values of $T$ and $K$ (Table \ref{tab:biasmsecomb}).

\begin{table}
\centering
\begin{tabular}{|c|c|c|c|c|c|c|c|c|c|}
\hline
\multirow{3}{*}{} & \multirow{3}{*}{True value} & \multicolumn{2}{|c|}{$T = 600$} & \multicolumn{2}{|c|}{$T = 1200$} & \multicolumn{2}{|c|}{$T = 600$} & \multicolumn{2}{|c|}{$T = 1200$} \\ 
& & \multicolumn{2}{|c|}{$K = 10$} & \multicolumn{2}{|c|}{$K = 10$} & \multicolumn{2}{|c|}{$K = 50$} & \multicolumn{2}{|c|}{$K = 50$} \\
\cline{3-10}
& & Bias & MSE & Bias & MSE & Bias & MSE & Bias & MSE \\
\hline
$\rho_{ns}$ & 0.75 & -0.17 & 0.03 & -0.18 & 0.04 & -0.14 & 0.03 & -0.17 & 0.04 \\
\hline
$\kappa$ & 0.75 & -0.21 & 0.05 & -0.21 & 0.05 & -0.19 & 0.04 & -0.19 & 0.04 \\
\hline
$\beta_0$ & 0.50 & 0.06 & 0.12 & 0.04 & 0.12 & 0.13 & 0.11 & 0.18 & 0.12 \\
\hline
$\beta_1$ & 7.5 & 1.3 & 10.1 & 1.6 & 13.4 & 0.0 & 2.6 & -0.7 & 1.7 \\
\hline
$\beta_2$ & -7.5 & -1.2 & 7.8 & -1.1 & 5.5 & 0.2 & 3.8 & 0.5 & 1.5 \\
\hline
$\beta_3$ & 0.0 & -0.1 & 5.0 & 0.4 & 4.3 & -0.1 & 2.5 & 0.0 & 1.7 \\
\hline
$\beta_d$ & 0.999 & -0.04 & 0.02 & -0.04 & 0.02 & -0.01 & 0.00 & -0.03 & 0.02 \\
\hline
$\mu$ & 500 & -8 & 795 & -13 & 543 & -22 & 7819 & -25 & 7277 \\
\hline
$\sigma$ & 25 & 2 & 341 & -3 & 32 & 1 & 197 & -1 & 217 \\
\hline
$\alpha$ & -1.78 & -0.45 & 2.64 & -0.28 & 1.30 & -0.14 & 1.81 & -0.67 & 2.17 \\
\hline
$\lambda$ & 0.85 & -0.02 & 0.001 & -0.02 & 0.001 & -0.02 & 0.002 & -0.03 & 0.001 \\
\hline
$\gamma$ & 0.90 & -0.03 & 0.001 & -0.03 & 0.001 & -0.03 & 0.001 & -0.03 & 0.001 \\
\hline
\end{tabular}
\caption{Estimates of bias and MSE for each parameter in the resource-memory model, averaged from 50 simulated movement tracks per treatment group. True values for each parameter are displayed on the left.}
\label{tab:biasmsecomb}
\end{table}

\subsection{Model application: grizzly bear case study}

According to our modelling framework, five of the eight grizzly bears exhibited consistently timed revisitations to previously visited locations in their home ranges (Table \ref{tab:bicbears}). When the data were broken up into one-year increments, model selection results varied annually, and sometimes differed even from the full dataset. For three of the bears (GF1008, GF1016, GM1046), the model identified as most explanatory of the bears' movement behaviors by BIC was different for the full dataset, the first subset, and the second subset. The resource-memory model was the most parsimonious explanation of the movement patterns of four bears, while the resource-only (2), memory-only (1), and null (1) models were also identified as most parsimonious in some cases. Four of the five memory-informed bears exhibited seasonal memory timescales close to one year ($\mu > 320$ days), while GF1016 had a $\mu$ value of 3 days. The six bears with resource selection included in their ``best model" displayed similar resource selection patterns: significant selection for areas indicative of berries and Arctic ground squirrels, avoidance of areas indicative of sweetvetch, and indifference to towns and cabins. When we considered the resources to be explicitly seasonal, the memory-only model was most commonly the ``best model" for the bears, with models including resources being much less common (Table \ref{tab:bicbears}).

\begin{table}[h]
\centering
\begin{tabular}{|c|c|c|c|c|}
\hline Bear ID & Full data & First subset & Second subset & Seasonal resources \\
\hline GF1004 & RM (38.54) & N (1.11) & RM (1.15) & RM (3.50) \\
\hline GF1008 & RM (17.53) & R (39.78) & N (4.90) & M (21.34) \\
\hline GF1016 & M (5.99) & N (18.24) & R (1.84) & M (22.32) \\
\hline GF1041 & RM (45.74) & R (18.24) & RM (24.62) & N (4.29) \\
\hline GF1086 & R (49.35) & R (5.45) & RM (19.68) & R (16.78) \\
\hline GF1107 & R (33.63) & R (30.17) & RM (83.54) & M (15.25) \\
\hline GF1130 & N (12.87) & RM (73.82) & N (17.13) & RM (16.71) \\
\hline GM1046 & RM (40.32) & R (6.03) & M (10.06) & M (19.71) \\
\hline
\end{tabular}
\caption{Model selection results for each bear in the Mackenzie Delta population. We list the hypothesis (N = null; R = resource-only; M = memory-only; RM = resource-memory) identified by BIC as most likely to be true given the data for the full dataset, the first subset, and the second subset. We also include results for the full dataset when resources were modelled as being explicitly seasonal. The numbers in parentheses are the difference in BIC between the best model and the second-best model.}
\label{tab:bicbears}
\end{table}

Our simulation study revealed that at smaller sample sizes, the model occasionally failed to identify memory from memory-informed simulated tracks, but this issue is remedied with double the data. An example was GF1008, where only 10 of the 100 simulated tracks were correctly identified as ``resource-memory" movements at $T = 600$. With $T = 1200$, this improved to 89. When we used likelihood ratio tests to compare the resource-memory model with the resource-only model (a special case of the resource-memory model) for GF1008, we found that at $T = 600$, 76 of our 100 simulated tracks registered a p-value below 0.05, indicating that the resource-memory model was significantly more explanatory than the resource-only model 76\% of the time. With $T = 1200$, this increased to 95. We observed similar trends for the other three resource-memory bears (GF1004, GF1041, GM1046) but not as strongly. It should be noted that GF1008 had the smallest estimate for $\beta_d$ (2.3) of these bears. When we performed BIC model selection on simulated tracks based on GF1086, a ``resource-only" bear, a false memory signal was identified more frequently with larger $T$ (from 4 to 12 out of 100). This trend was not replicated for GF1107, the other ``resource-only" bear (decrease from 10 to 7).

\section{Discussion}

Our model builds on existing literature to identify unique behavioral and cognitive mechanisms from animal movement data. Using advanced computational techniques, this novel and complex modelling framework can provide statistical inference for a variety of ecological systems. Our simulation studies provided insight on the viability of the model for different amounts of data.

We formulated a model that expresses parameters with clear biological implications to aid in the interpretation of our results, but we had to do so carefully to ensure that these parameters could be estimated accurately. Finding a set of biologically meaningful parameters with low mean squared error (Table \ref{tab:biasmsecomb}) required a degree of trial and error, especially for $\beta_0$ and $\beta_d$. We chose to express them in a way that makes sense both biologically (where they represent relative probabilities) and mathematically (where they can easily be estimated with less error). While we can redefine these parameters without actually changing our likelihood function, we made sure to define parameters that are easy to estimate and biologically meaningful.

Our results provided support for a positive effect of the amount of location data and control points on parameter estimation, with the number of control points having a negligible effect on model selection accuracy. However, at all treatment groups, parameter estimates were occasionally inaccurate (Fig. \ref{fig:beta1violin}, Table \ref{tab:biasmsecomb}), and the model occasionally mistook the movement mechanisms driving the simulated tracks (Table \ref{tab:biccountssims}). These outliers may be due to the stochasticity of simulated movement tracks; for example, a resource-only simulated animal may happen to visit similar portions of its landscape at a coincidentally regular interval, which the model might mistake for memory-informed movement. Conversely, an animal following the ``memory-only" rules may coincidentally visit locations that happen to be particularly high (or low) in specific resource values, resulting in the movement track being best explained by the resource-memory or even the resource-only model.

Increasing the number of observed animal locations ($T$) improved our results, but we are more encouraged by the positive effects of simulating additional control points ($K$). While increasing $T$ may require such costly tasks as using longer-lasting tracking devices, re-capturing animals and equipping them with new tracking devices periodically, or increasing the temporal resolution of tracking devices, increasing $K$ is easy to do post-hoc. While increasing $K$ may not yield benefits as large as increasing $T$, the cost of increasing $K$ is much smaller.

Our simulated tracks consistently underestimated $\rho_{ns}$ and $\kappa$ in the resource-only and resource-memory models (Table \ref{tab:biasmsecomb}), which is  an artifact of the way we simulated the data. In these models, the animal ``chooses" a step from $N_r$ proposed steps, which are simulated from $k$, which depends on $\rho_{ns}$ and $\kappa$. Our simulated landscapes are spatially autocorrelated, so if the simulated animal found itself in a resource-rich patch, it would be very likely to stay put. These movements are also less directionally autocorrelated than would be suggested by $\kappa$ for similar reasons. Using an integrated step selection function \citep{Avgar2016} could remedy these issues but for our purposes, it adds additional complexity to the model and is not our primary concern.

Our estimates of bias and MSE for $\alpha$ did not consistently decrease with increases in the amount of location data or the number of control points, potentially because of an odd bimodal distribution of parameter estimates (Appendix: Fig. S1). The larger portion of this bimodal distribution is clearly centered around the true value of approximately -1.78 for all four treatment groups, but curiously the ``second" mode, which appears to be centered around -4.5, seems to account for more of the estimates $T$ and $K$ increase. These smaller estimates for $\alpha$ would imply that the hypothetical organisms moving according to our simulation rules occasionally behave with a much wider understanding of their environment, which they perceive to be spatially heterogeneous. The exact cause of these patterns requires further investigation.

When we applied the analysis to field data, we notice that the model's effectiveness, especially when it comes to identifying a memory signal, increased greatly with sample size. Our simulations revealed that the model may miss a memory signal with inadequate data, which could explain the disparity between subsets of the data in Table \ref{tab:bicbears}. It must also be noted that, as stated in the Introduction, the goal of this model is not to determine whether or not grizzly bears have spatial memory; we are more interested in if they use that memory the way we have hypothesized. If the resource-only model is the ``best model" for a bear, it may just mean that they are using memory in some other way. While it is possible that grizzly bears, especially females that take on different reproductive roles in different years, would change their movement strategies between years, it is also likely that the model may have not had enough data to identify a memory signal in an individual year. With twice data, the simulations accurately identified memory more often, suggesting that the memory signals identified for the entirety of each bear's track are legitimate. Nevertheless, even with these considerations, we see that half the bears in the dataset exhibited patterns following the resource-memory hypothesis, suggesting a strong influence of both habitat selection as well as spatial memory on the movement of grizzly bears in the Mackenzie Delta.

We occasionally observed ``false positives" (the model identified memory as a driver of movement from memoryless simulated data) that increased in longer animal tracks ($K = 1200$ vs. $K = 600$) when simulating tracks with the grizzly bear data. This trend may be an artifact of how the Mackenzie Delta landscape data influenced simulated tracks, since false positives were much less frequent in the simulation study with artificial landscapes. When comparing this result with the real-life subsetting for the bears, we saw examples of subsetted data registering memory when the full data set did not, but we also saw examples of the opposite.

Our modelling framework operates under the assumption that resources vary in time, forcing animals to exhibit seasonal movement patterns within their home ranges. We handled this assumption in two different ways: by explicitly defining this temporal variation, and by indirectly incorporating it into the resource-memory and memory-only models. In this case, explicitly defining the seasonality of the grizzly bear resources made the memory-only model (which is primarily meant for situations when sufficient environmental data may not be available) much more effective. We suggest that making arbitrary assumptions about these timings may not always improve model parsimony, and instead may overshadow patterns and behaviors we are interested in. An alternative method to capture this variation would be to assume that $\mu$ is informative about how long resources take to re-appear, and as a result how long animals take to return to them. 

Due to the novelty of this contribution, we accept that there will be opportunities to build on and improve the approach. Particularly interesting is the addition of more behavioral states to the model. We applied a hidden Markov component to the model mainly to avoid mistaking stationary periods for recursive movement on a short timescale, but adding many states (e.g., a memoryless searching state and a memory-informed navigating state) could provide insight on the frequency of these movements. One such adjustment could involve changing the form of $W_M$ and $W_{RM}$ such that they are truly periodic; this could be done by changing $\varphi$ from a traditional Gaussian to a wrapped Gaussian. Making this change would imply that animals are influenced to revisit locations they visited $k \mu$ time steps ago for all positive integers $k$. Including such a mechanism would also potentially warrant the incorporation of explicit memory decay, which we omitted but could be useful when longer timescales or wrapped distributions are involved. Revising $\varphi$ to a mixture of multiple Gaussians instead could also be used to test the hypothesis that animals perform recursive movements on different, asymmetrical scales. Modifying the formulation of the cognitive map $Z$ (e.g., to something resembling a discrete-time analog of the territory interaction model from \citealp{Potts2016}) could also be an opportunity to improve and tweak the model. Connections to the work of \cite{Potts2016} could also be made by incorporating territoriality or the presence of other individuals into the model somehow, potentially as a ``resource" covariate. A final point for future work would be to redevelop this model from the perspective of integrated step selection analysis (iSSA; \citealp{Avgar2016}). Here, we could analyze how animal movement behavior is directly influenced by covariates such as its distance from previously visited locations or the strength of its reliance on spatial memory.

While we used grizzly bears as a case study, the model was designed to be general and can be applied to a variety of different systems. Many animals, including turkey vultures (\textit{Cathartes aura}; \citealp{Holland2017}), black vultures (\textit{Coragyps atratus}; \citealp{Holland2017}), caribou (\textit{Rangifer tarandus}; \citealp{Lafontaine2017}), and eastern indigo snakes (\textit{Drymarchon couperi}; \citealp{Bauder2016}), perform seasonal movements within their home ranges. For data with higher temporal resolution, it would be possible to model complex time-dependent recursive movements on a diel scale, since many animals exhibit repetitive day-to-day movements within their home range \citep{Christiansen2016, Herbig2016}. Collecting data at finer temporal resolutions would be beneficial for inference on memory-informed movement, assuming observation errors are accounted for. Even patrolling predators, which were modelled by \cite{Schlagel2014}, could be modelled using our framework, although we may expect estimates for $\mu$ to be smaller than in the grizzly bears. \cite{Schlagel2017} displayed the importance of time since last visit for gray wolves, but insight on when exactly wolves deem parts of their home range ``re-visitable" could be interesting. Of course, migration is also seasonal and predictable, and although it is typically difficult to obtain environmental data for an animal's entire migratory route, spatial memory has been identified as a key driver of migration in many instances \citep{Mueller2008, Mueller2011, Fagan2013, Bracis2017, Merkle2019}. Fitting this model to migratory populations could provide insights on how to quantify or potentially even predict these mechanisms.

\section{Conclusions}

Our model uses patterns in animal movement data to obtain information on complex time-dependent spatial memory patterns. Made possible by advanced computational techniques, we expand on existing literature from animal movement modelling as well as animal cognition to generate a model that can be applied to a variety of ecological systems. The model can estimate the timing of recursive movement patterns observed in an animal, which is novel, and also allows for the interaction of present-time resource selection and memory-informed navigation. We verify our model fitting process using simulated data before testing its utility on GPS collar data from grizzly bears, finding that this very complex model can be effective without need for immense data collection. We hope to apply this model more broadly to animals with different foraging strategies as a means to compare the nature of time-dependent memory mechanisms in different ecological systems.

\section{Acknowledgements}

PRT was funded by the Ashley and Janet Cameron Graduate Scholarship, with support from UAlberta North, as well as the Alberta Graduate Excellence Scholarship. PRT was also supported by a University of Alberta Master's Recruitment Award as well as a University of Alberta Doctoral Recruitment Award. MAL gratefully acknowledges support from a Canada Research Chair and NSERC Discovery grant. The authors declare no conflict of interest.

\section{Conflict of interest}

We have no conflict of interest to declare.

\section{Author contributions}

PRT designed the analysis with recommendations from MAL, AED, and MAE. MAE and AED collected and supplied grizzly bear data. PRT wrote the draft of the manuscript. All authors have reviewed and provided modifications to the manuscript.

\section{Data accessibility}

We have uploaded all grizzly bear movement data and code necessary for conducting the analysis to a GitHub repository that is available through Zenodo \citep{Data2021}.


\newpage

\begin{thebibliography}{}

\bibitem[Albertsen et~al., 2015]{Albertsen2015}
Albertsen, C.~M., Whoriskey, K., Yurkowski, D., Nielsen, A., and Flemming,
  J.~M. (2015).
\newblock Fast fitting of non-gaussian state-space models to animal movement
  data via template model builder.
\newblock {\em Ecology}, 96:2598--2604.

\bibitem[Allen and Fortin, 2013]{Allen2013}
Allen, T.~A. and Fortin, N.~J. (2013).
\newblock The evolution of episodic memory.
\newblock {\em Proceedings of the National Academy of Sciences},
  110(Supplement_2):10379–10386.

\bibitem[Asensio et~al., 2011]{Asensio2011}
Asensio, N., Brockelman, W.~Y., Malaivijitnond, S., and Reichard, U.~H. (2011).
\newblock Gibbon travel paths are goal oriented.
\newblock {\em Animal Cognition}, 14(3):395--405.

\bibitem[Auger-Méthé et~al., 2017]{Auger2017}
Auger-Méthé, M., Albertsen, C.~M., Jonsen, I.~D., Derocher, A.~E., Lidgard,
  D.~C., Studholme, K.~R., Bowen, W.~D., Crossin, G.~T., and Mills~Flemming, J.
  (2017).
\newblock Spatiotemporal modelling of marine movement data using template model
  builder (tmb).
\newblock {\em Marine Ecology Progress Series}, 565:237--249.

\bibitem[Auger-Méthé et~al., 2015]{Auger2015}
Auger-Méthé, M., Derocher, A.~E., Plank, M.~J., Codling, E.~A., Lewis, M.~A.,
  and Börger, L. (2015).
\newblock Differentiating the {Lévy} walk from a composite correlated random
  walk.
\newblock {\em Methods in Ecology and Evolution}, 6(10):1179--1189.

\bibitem[Avgar et~al., 2015]{Avgar2015}
Avgar, T., Baker, J.~A., Brown, G.~S., Hagens, J.~S., Kittle, A.~M., Mallon,
  E.~E., McGreer, M.~T., Mosser, A., Newmaster, S.~G., Patterson, B.~R., and
  et~al. (2015).
\newblock Space-use behaviour of woodland caribou based on a cognitive movement
  model.
\newblock {\em Journal of Animal Ecology}, 84(4):1059–1070.

\bibitem[Avgar et~al., 2013]{Avgar2013}
Avgar, T., Deardon, R., and Fryxell, J.~M. (2013).
\newblock An empirically parameterized individual based model of animal
  movement, perception, and memory.
\newblock {\em Ecological Modelling}, 251:158--172.

\bibitem[Avgar et~al., 2016]{Avgar2016}
Avgar, T., Potts, J.~R., Lewis, M.~A., and Boyce, M.~S. (2016).
\newblock Integrated step selection analysis: bridging the gap between resource
  selection and animal movement.
\newblock {\em Methods in Ecology and Evolution}, 7(5):619--630.

\bibitem[Barker and Derocher, 2009]{Barker2009}
Barker, O.~E. and Derocher, A.~E. (2009).
\newblock Brown bear (\textit{{Ursus} arctos}) predation of broad whitefish
  (\textit{{Coregonus} nasus}) in the mackenzie delta region, northwest
  territories.
\newblock {\em Arctic}, 62(3):312--316.

\bibitem[Barker and Derocher, 2010]{Barker2010}
Barker, O.~E. and Derocher, A.~E. (2010).
\newblock Habitat selection by arctic ground squirrels (\textit{{Spermophilus}
  parryii}).
\newblock {\em Journal of Mammalogy}, 91(5):1251--1260.

\bibitem[Bauder et~al., 2016]{Bauder2016}
Bauder, J.~M., Breininger, D.~R., Bolt, M.~R., Legare, M.~L., Jenkins, C.~L.,
  Rothermel, B.~B., and McGarigal, K. (2016).
\newblock Seasonal variation in eastern indigo snake (\textit{{Drymarchon}
  couperi}) movement patterns and space use in peninsular florida at multiple
  temporal scales.
\newblock {\em Herpetologica}, 72(3):214--226.

\bibitem[Bennett, 1996]{Bennett1996}
Bennett, A.~T. (1996).
\newblock Do animals have cognitive maps?
\newblock {\em Journal of Experimental Biology}, 199:219--224.

\bibitem[Berger-Tal and Bar-David, 2015]{Berger2015}
Berger-Tal, O. and Bar-David, S. (2015).
\newblock Recursive movement patterns: review and synthesis across species.
\newblock {\em Ecosphere}, 6(9).

\bibitem[Boyce and McDonald, 1999]{Boyce1999}
Boyce, M.~S. and McDonald, L. (1999).
\newblock Relating populations to habitats using resource selection functions.
\newblock {\em TREE}, 14(7):268--272.

\bibitem[Bracis and Mueller, 2017]{Bracis2017}
Bracis, C. and Mueller, T. (2017).
\newblock Memory, not just perception, plays an important role in terrestrial
  mammalian migration.
\newblock {\em Proc Biol Sci}, 284(1855).

\bibitem[Buck and Barnes, 1999]{Buck1999}
Buck, C.~L. and Barnes, B.~M. (1999).
\newblock Annual cycle of body composition and hibernation in free-living
  arctic ground squirrels.
\newblock {\em Journal of Mammalogy}, 80(2):430--442.

\bibitem[Buderman et~al., 2018]{Buderman2018}
Buderman, F.~E., Hooten, M.~B., Alldredge, M.~W., Hanks, E.~M., and Ivan, J.~S.
  (2018).
\newblock Time-varying predatory behavior is primary predictor of fine-scale
  movement of wildland-urban cougars.
\newblock {\em Mov Ecol}, 6:22.

\bibitem[Burnham and Anderson, 2004]{Burnham2004}
Burnham, K.~P. and Anderson, D.~R. (2004).
\newblock Multimodel inference.
\newblock {\em Sociological Methods \& Research}, 33(2):261--304.

\bibitem[Charnov, 1976]{Charnov1976}
Charnov, E. (1976).
\newblock Optimal foraging, the marginal value theorem.
\newblock {\em Theoretical Population Biology}, 9:129--136.

\bibitem[Christiansen et~al., 2016]{Christiansen2016}
Christiansen, F., Esteban, N., Mortimer, J.~A., Dujon, A.~M., and Hays, G.~C.
  (2016).
\newblock Diel and seasonal patterns in activity and home range size of green
  turtles on their foraging grounds revealed by extended fastloc-gps tracking.
\newblock {\em Marine Biology}, 164(1).

\bibitem[Clayton and Dickinson, 1998]{Clayton1998}
Clayton, N.~S. and Dickinson, A. (1998).
\newblock Episodic-like memory during cache recovery by scrub jays.
\newblock {\em Nature}, 395(6699):272--274.

\bibitem[Codling et~al., 2008]{Codling2008}
Codling, E.~A., Plank, M.~J., and Benhamou, S. (2008).
\newblock Random walk models in biology.
\newblock {\em J R Soc Interface}, 5(25):813--34.

\bibitem[Crystal, 2018]{Crystal2018}
Crystal, J.~D. (2018).
\newblock Animal models of episodic memory.
\newblock {\em Comparative Cognition \& Behavior Reviews}, 13:105–122.

\bibitem[Dalziel et~al., 2008]{Dalziel2008}
Dalziel, B.~D., Morales, J.~M., and Fryxell, J.~M. (2008).
\newblock Fitting probability distributions to animal movement trajectories:
  using artificial neural networks to link distance, resources, and memory.
\newblock {\em Am Nat}, 172(2):248--58.

\bibitem[Davies and Houston, 1981]{Davies1981}
Davies, N. and Houston, A. (1981).
\newblock Owners and satellites: The economics of territory defence in the pied
  wagtail, \textit{{Motacilla} alba}.
\newblock {\em Journal of Animal Ecology}, 50(1):157--180.

\bibitem[Duchesne et~al., 2015]{Duchesne2015}
Duchesne, T., Fortin, D., and Rivest, L.~P. (2015).
\newblock Equivalence between step selection functions and biased correlated
  random walks for statistical inference on animal movement.
\newblock {\em PLoS One}, 10(4):e0122947.

\bibitem[Eacott and Easton, 2010]{Eacott2010}
Eacott, M.~J. and Easton, A. (2010).
\newblock Episodic memory in animals: Remembering which occasion.
\newblock {\em Neuropsychologia}, 48(8):2273–2280.

\bibitem[Edwards et~al., 2011]{Edwards2011}
Edwards, M., Derocher, A.~E., Hobson, K., Branigan, M., and Nagy, J. (2011).
\newblock Fast carnivores and slow herbivores: differential foraging strategies
  among grizzly bears in the canadian arctic.
\newblock {\em Oecologia}, 165:877--889.

\bibitem[Edwards and Derocher, 2015]{Edwards2015}
Edwards, M.~A. and Derocher, A.~E. (2015).
\newblock Mating-related behaviour of grizzly bears inhabiting marginal habitat
  at the periphery of their north american range.
\newblock {\em Behavioural Processes}, 111:75--83.

\bibitem[Edwards et~al., 2009]{Edwards2009}
Edwards, M.~A., Nagy, J.~A., and Derocher, A.~E. (2009).
\newblock Low site fidelity and home range drift in a wide-ranging, large
  arctic omnivore.
\newblock {\em Animal Behaviour}, 77(1):23--28.

\bibitem[Fagan et~al., 2013]{Fagan2013}
Fagan, W.~F., Lewis, M.~A., Auger-Methe, M., Avgar, T., Benhamou, S., Breed,
  G., LaDage, L., Schlagel, U.~E., Tang, W.~W., Papastamatiou, Y.~P., Forester,
  J., and Mueller, T. (2013).
\newblock Spatial memory and animal movement.
\newblock {\em Ecology Letters}, 16(10):1316--29.

\bibitem[Farnsworth and Beecham, 1999]{Farnsworth1999}
Farnsworth, K.~D. and Beecham, J.~A. (1999).
\newblock How do grazers achieve their distribution? a continuum of models from
  random diffusion to the ideal free distribution using biased random walks.
\newblock {\em The American Naturalist}, 153(5):509–526.

\bibitem[Fischer and Lewis, 2020]{Fischer2020}
Fischer, S.~M. and Lewis, M.~A. (2020).
\newblock A robust and efficient algorithm to find profile likelihood
  confidence intervals.
\newblock {\em ArXiv}, pages 1--18.

\bibitem[Forester et~al., 2009]{Forester2009}
Forester, J.~D., Im, H.~K., and Rathouz, P.~J. (2009).
\newblock Accounting for animal movement in estimation of resource selection
  functions: sampling and data analysis.
\newblock {\em Ecology}, 90(12):3554–3565.

\bibitem[Fortin et~al., 2005]{Fortin2005}
Fortin, D., Beyer, H., Boyce, M.~S., Smith, D., Duchesne, T., and Mao, J.
  (2005).
\newblock Wolves influence elk movements: Behavior shapes a trophic cascade in
  yellowstone national park.
\newblock {\em Ecology}, 86(5):1320--1330.

\bibitem[Gaillard et~al., 2010]{Gaillard2010}
Gaillard, J.~M., Hebblewhite, M., Loison, A., Fuller, M., Powell, R., Basille,
  M., and Van~Moorter, B. (2010).
\newblock Habitat-performance relationships: finding the right metric at a
  given spatial scale.
\newblock {\em Philos Trans R Soc Lond B Biol Sci}, 365(1550):2255--65.

\bibitem[Gau et~al., 2002]{Gau2002}
Gau, R.~J., Case, R.~L., Fenner, D., and McLoughlin, P.~D. (2002).
\newblock Feeding patterns of barren-ground grizzly bears in the central
  canadian arctic.
\newblock {\em Arctic}, 55(4):339--344.

\bibitem[Gerber et~al., 2019]{Gerber2019}
Gerber, B.~D., Hooten, M.~B., Peck, C.~P., Rice, M.~B., Gammonley, J.~H., Apa,
  A.~D., Davis, A.~J., and Lemaître, J. (2019).
\newblock Extreme site fidelity as an optimal strategy in an unpredictable and
  homogeneous environment.
\newblock {\em Functional Ecology}, 33(9):1695--1707.

\bibitem[Graham et~al., 2012]{Graham2012}
Graham, R.~T., Witt, M.~J., Castellanos, D.~W., Remolina, F., Maxwell, S.,
  Godley, B.~J., and Hawkes, L.~A. (2012).
\newblock Satellite tracking of manta rays highlights challenges to their
  conservation.
\newblock {\em PLoS One}, 7(5):e36834.

\bibitem[Harel and Nathan, 2018]{Harel2018}
Harel, R. and Nathan, R. (2018).
\newblock The characteristic time-scale of perceived information for
  decision-making: Departure from thermal columns in soaring birds.
\newblock {\em Functional Ecology}, 32(8):2065--2072.

\bibitem[Herbig and Szedlmayer, 2016]{Herbig2016}
Herbig, J.~L. and Szedlmayer, S.~T. (2016).
\newblock Movement patterns of gray triggerfish, \textit{{Balistes} capriscus},
  around artificial reefs in the northern gulf of mexico.
\newblock {\em Fisheries Management and Ecology}, 23(5):418--427.

\bibitem[Holland et~al., 2017]{Holland2017}
Holland, A.~E., Byrne, M.~E., Bryan, A.~L., DeVault, T.~L., Rhodes, O.~E., and
  Beasley, J.~C. (2017).
\newblock Fine-scale assessment of home ranges and activity patterns for
  resident black vultures (\textit{{Coragyps} atratus}) and turkey vultures
  (\textit{{Cathartes} aura}).
\newblock {\em PLoS One}, 12(7):e0179819.

\bibitem[Janmaat et~al., 2006]{Janmaat2006}
Janmaat, K.~R., Byrne, R.~W., and Zuberbuhler, K. (2006).
\newblock Primates take weather into account when searching for fruits.
\newblock {\em Curr Biol}, 16(12):1232--7.

\bibitem[Johnson, 1980]{Johnson1980}
Johnson, D. (1980).
\newblock The comparison of usage and availability measurements for evaluating
  resource preference.
\newblock {\em Ecology}, 61(1):65--71.

\bibitem[Jonsen et~al., 2013]{Jonsen2013}
Jonsen, I.~D., Basson, M., Bestley, S., Bravington, M.~V., Patterson, T.~A.,
  Pedersen, M.~W., Thomson, R., Thygesen, U.~H., and Wotherspoon, S.~J. (2013).
\newblock State-space models for bio-loggers: A methodological road map.
\newblock {\em Deep Sea Research Part II: Topical Studies in Oceanography},
  88-89:34--46.

\bibitem[Kays et~al., 2015]{Kays2015}
Kays, R., Crofoot, M.~C., Jetz, W., and Wikelski, M. (2015).
\newblock Terrestrial animal tracking as an eye on life and planet.
\newblock {\em Science}, 348(6240):aaa2478.

\bibitem[Kristensen et~al., 2016]{Kristensen2016}
Kristensen, K., Nielsen, A., Berg, C.~W., Skaug, H., and Bell, B.~M. (2016).
\newblock {TMB}: Automatic differentiation and {Laplace} approximation.
\newblock {\em Journal of Statistical Software}, 70(5).

\bibitem[Lafontaine et~al., 2017]{Lafontaine2017}
Lafontaine, A., Drapeau, P., Fortin, D., and St-Laurent, M.~H. (2017).
\newblock Many places called home: the adaptive value of seasonal adjustments
  in range fidelity.
\newblock {\em J Anim Ecol}, 86(3):624--633.

\bibitem[Macdonald et~al., 1995]{Macdonald1995}
Macdonald, R.~W., Paton, D.~W., Carmack, E.~C., and Omstedt, A. (1995).
\newblock The freshwater budget and under-ice spreading of mackenzie river
  water in the canadian beaufort sea based on salinity and18o/16o measurements
  in water and ice.
\newblock {\em Journal of Geophysical Research}, 100(C1).

\bibitem[MacHutchon and Wellwood, 2003]{MacHutchon2003}
MacHutchon, A.~G. and Wellwood, D.~W. (2003).
\newblock Grizzly bear food habits in the northern yukon, canada.
\newblock {\em Ursus}, 14(2):225--235.

\bibitem[Marchand et~al., 2017]{Marchand2017}
Marchand, P., Garel, M., Bourgoin, G., Duparc, A., Dubray, D., Maillard, D.,
  and Loison, A. (2017).
\newblock Combining familiarity and landscape features helps break down the
  barriers between movements and home ranges in a non-territorial large
  herbivore.
\newblock {\em J Anim Ecol}, 86(2):371--383.

\bibitem[Martin-Ordas et~al., 2010]{Martin2010}
Martin-Ordas, G., Haun, D., Colmenares, F., and Call, J. (2010).
\newblock Keeping track of time: evidence for episodic-like memory in great
  apes.
\newblock {\em Anim Cogn}, 13(2):331--40.

\bibitem[Mauritzen et~al., 2001]{Mauritzen2001}
Mauritzen, M., Derocher, A.~E., and Wiig, {\O}. (2001).
\newblock Space-use strategies of female polar bears in a dynamic sea ice
  habitat.
\newblock {\em Canadian Journal of Zoology}, 79(9):1704--1713.

\bibitem[Merkle et~al., 2014]{Merkle2014}
Merkle, J.~A., Fortin, D., and Morales, J.~M. (2014).
\newblock A memory-based foraging tactic reveals an adaptive mechanism for
  restricted space use.
\newblock {\em Ecol Lett}, 17(8):924--31.

\bibitem[Merkle et~al., 2019]{Merkle2019}
Merkle, J.~A., Sawyer, H., Monteith, K.~L., Dwinnell, S. P.~H., Fralick, G.~L.,
  and Kauffman, M.~J. (2019).
\newblock Spatial memory shapes migration and its benefits: evidence from a
  large herbivore.
\newblock {\em Ecol Lett}, 22(11):1797--1805.

\bibitem[Morales et~al., 2004]{Morales2004}
Morales, J.~M., Haydon, D.~T., Frair, J., Holsinger, K.~E., and Fryxell, J.~M.
  (2004).
\newblock Extracting more out of relocation data: Building movement models as
  mixtures of random walks.
\newblock {\em Ecology}, 85(9):2436--2445.

\bibitem[Morey et~al., 2007]{Morey2007}
Morey, P.~S., Gese, E.~M., and Gehrt, S.~D. (2007).
\newblock Spatial and temporal variation in the diet of coyotes in the
  {Chicago} metropolitan area.
\newblock {\em The American Midland Naturalist}, 158(1):147--161.

\bibitem[Mueller and Fagan, 2008]{Mueller2008}
Mueller, T. and Fagan, W.~F. (2008).
\newblock Search and navigation in dynamic environments - from individual
  behaviors to population distributions.
\newblock {\em Oikos}, 117:654--664.

\bibitem[Mueller et~al., 2011]{Mueller2011}
Mueller, T., Fagan, W.~F., and Grimm, V. (2011).
\newblock Integrating individual search and navigation behaviors in mechanistic
  movement models.
\newblock {\em Theoretical Ecology}, 4(3):341--355.

\bibitem[Munoz-Lopez and Morris, 2009]{Munoz2009}
Munoz-Lopez, M. and Morris, R. (2009).
\newblock Episodic memory: Assessment in animals.
\newblock {\em Encyclopedia of Neuroscience}, 3:1173–1182.
\newblock journalAbbreviation: Encyclopedia of Neuroscience.

\bibitem[Nathan et~al., 2008]{Nathan2008}
Nathan, R., Getz, W., Revilla, E., Holyoak, M., Kadmon, R., Saltz, D., and
  Smouse, P.~E. (2008).
\newblock A movement ecology paradigm for unifying organismal movement
  research.
\newblock {\em PNAS}, 105(49):19052--19059.

\bibitem[Normand and Boesch, 2009]{Normand2009}
Normand, E. and Boesch, C. (2009).
\newblock Sophisticated {Euclidean} maps in forest chimpanzees.
\newblock {\em Animal Behaviour}, 77(5):1195--1201.

\bibitem[O'Keefe and Nadel, 1978]{OKeefe1978}
O'Keefe, J. and Nadel, L. (1978).
\newblock {\em The hippocampus as a cognitive map}.
\newblock Clarendon Press, Oxford, United Kingdom.

\bibitem[Oliveira-Santos et~al., 2016]{OliveiraSantos2016}
Oliveira-Santos, L.~G., Forester, J.~D., Piovezan, U., Tomas, W.~M., and
  Fernandez, F.~A. (2016).
\newblock Incorporating animal spatial memory in step selection functions.
\newblock {\em J Anim Ecol}, 85(2):516--24.

\bibitem[Pan and Wu, 1998]{Pan1998}
Pan, L. and Wu, L. (1998).
\newblock A hybrid global optimization method for inverse estimation of
  hydraulic parameters: Annealing-simplex method.
\newblock {\em Water Resources Research}, 34(9):2261--2269.

\bibitem[Papastamatiou et~al., 2013]{Papastamatiou2013}
Papastamatiou, Y.~P., Meyer, C.~G., Carvalho, F., Dale, J.~J., Hutchinson,
  M.~R., and Holland, K.~N. (2013).
\newblock Telemetry and random-walk models reveal complex patterns of partial
  migration in a large marine predator.
\newblock {\em Ecology}, 94(11):2595--2606.

\bibitem[Parzen, 1961]{Parzen1961}
Parzen, E. (1961).
\newblock On estimation of a probability density function and mode.
\newblock {\em The Annals of Mathematical Statistics}, 33:1065--1076.

\bibitem[Potts and Lewis, 2016]{Potts2016}
Potts, J.~R. and Lewis, M.~A. (2016).
\newblock How memory of direct animal interactions can lead to territorial
  pattern formation.
\newblock {\em Journal of The Royal Society Interface}, 13(118):20160059.

\bibitem[Prokopenko et~al., 2017]{Prokopenko2017}
Prokopenko, C.~M., Boyce, M.~S., Avgar, T., and Tulloch, A. (2017).
\newblock Characterizing wildlife behavioural responses to roads using
  integrated step selection analysis.
\newblock {\em Journal of Applied Ecology}, 54(2):470--479.

\bibitem[Richter et~al., 2020]{Richter2020}
Richter, L., Balkenhol, N., Raab, C., Reinecke, H., Meißner, M., Herzog, S.,
  Isselstein, J., and Signer, J. (2020).
\newblock So close and yet so different: the importance of considering temporal
  dynamics to understand habitat selection.
\newblock {\em Basic and Applied Ecology}, 43:99--109.

\bibitem[Riotte-Lambert et~al., 2015]{Riotte2015}
Riotte-Lambert, L., Benhamou, S., and Chamaille-Jammes, S. (2015).
\newblock How memory-based movement leads to nonterritorial spatial
  segregation.
\newblock {\em Am Nat}, 185(4):E103--16.

\bibitem[Riotte-Lambert et~al., 2017]{Riotte2017}
Riotte-Lambert, L., Benhamou, S., and Chamaillé-Jammes, S. (2017).
\newblock From randomness to traplining: a framework for the study of routine
  movement behavior.
\newblock {\em Behavioral Ecology}, 28(1):280–287.

\bibitem[Schlägel and Lewis, 2014]{Schlagel2014}
Schlägel, U.~E. and Lewis, M.~A. (2014).
\newblock Detecting effects of spatial memory and dynamic information on animal
  movement decisions.
\newblock {\em Methods in Ecology and Evolution}, 5(11):1236--1246.

\bibitem[Schlägel et~al., 2017]{Schlagel2017}
Schlägel, U.~E., Merrill, E.~H., and Lewis, M.~A. (2017).
\newblock Territory surveillance and prey management: Wolves keep track of
  space and time.
\newblock {\em Ecol Evol}, 7(20):8388--8405.

\bibitem[Sciaini et~al., 2018]{Sciaini2018}
Sciaini, M., Fritsch, M., Scherer, C., Simpkins, C.~E., and Golding, N. (2018).
\newblock Nlmr and landscapetools: An integrated environment for simulating and
  modifying neutral landscape models in {R}.
\newblock {\em Methods in Ecology and Evolution}, 9(11):2240--2248.

\bibitem[Shevtsova et~al., 1995]{Shevtsova1995}
Shevtsova, A., Ojala, A., Neuvonen, S., Vieno, M., and Haukioja, E. (1995).
\newblock Growth and reproduction of dwarf shrubs in a subarctic plant
  community: Annual variation and above-ground interactions with neighbours.
\newblock {\em Journal of Ecology}, 83(2):263--275.

\bibitem[Siniff and Jessen, 1969]{Siniff1969}
Siniff, D. and Jessen, C. (1969).
\newblock A simulation model of animal movement patterns.
\newblock {\em Advances in Ecological Research}, 6:185--219.

\bibitem[Skaug and Fournier, 2006]{Skaug2006}
Skaug, H.~J. and Fournier, D.~A. (2006).
\newblock Automatic approximation of the marginal likelihood in non-gaussian
  hierarchical models.
\newblock {\em Computational Statistics \& Data Analysis}, 51(2):699--709.

\bibitem[Sturz et~al., 2006]{Sturz2006}
Sturz, B.~R., Bodily, K.~D., and Katz, J.~S. (2006).
\newblock Evidence against integration of spatial maps in humans.
\newblock {\em Anim Cogn}, 9(3):207--17.

\bibitem[Thompson et~al., 2021]{Data2021}
Thompson, P.~R., Derocher, A.~E., Edwards, M.~A., and Lewis, M.~A. (2021).
\newblock {pthompson234/detectingmemory: Detecting seasonal episodic-like
  spatiotemporal memory patterns using animal movement modelling: Updates to
  supplementary information. https://doi.org/10.5281/zenodo.5557536}.

\bibitem[Thurfjell et~al., 2014]{Thurfjell2014}
Thurfjell, H., Ciuti, S., and Boyce, M.~S. (2014).
\newblock Applications of step-selection functions in ecology and conservation.
\newblock {\em Movement Ecology}, 2(4):1--12.

\bibitem[Tolman, 1948]{Tolman1948}
Tolman, E.~C. (1948).
\newblock Cognitive maps in rats and men.
\newblock {\em The Psychological Review}, 55(4):189--208.

\bibitem[Van~Moorter et~al., 2009]{VanMoorter2009}
Van~Moorter, B., Visscher, D., Benhamou, S., Börger, L., Boyce, M.~S., and
  Gaillard, J.-M. (2009).
\newblock Memory keeps you at home: a mechanistic model for home range
  emergence.
\newblock {\em Oikos}, 118(5):641--652.

\bibitem[Vergara et~al., 2016]{Vergara2016}
Vergara, P.~M., Soto, G.~E., Moreira-Arce, D., Rodewald, A.~D., Meneses, L.~O.,
  and Perez-Hernandez, C.~G. (2016).
\newblock Foraging behaviour in {Magellanic} woodpeckers is consistent with a
  multi-scale assessment of tree quality.
\newblock {\em PLoS One}, 11(7):e0159096.

\bibitem[Whoriskey et~al., 2017]{Whoriskey2017}
Whoriskey, K., Auger-Methe, M., Albertsen, C.~M., Whoriskey, F.~G., Binder,
  T.~R., Krueger, C.~C., and Mills~Flemming, J. (2017).
\newblock A hidden {Markov} movement model for rapidly identifying behavioral
  states from animal tracks.
\newblock {\em Ecol Evol}, 7(7):2112--2121.

\bibitem[Wiktander et~al., 2001]{Wiktander2001}
Wiktander, U., Olsson, O., and Nilsson, S.~J. (2001).
\newblock Seasonal variation in home-range size, and habitat area requirement
  of the lesser spotted woodpecker (\textit{{Dendrocopos} minor}) in southern
  sweden.
\newblock {\em Biological Conservation}, 100:387--385.

\bibitem[Worton, 1989]{Worton1989}
Worton, B. (1989).
\newblock Kernel methods for estimating the utilization distribution in
  home-range studies.
\newblock {\em Ecology}, 70(1):164--168.

\end{thebibliography}
\end{document}